\documentclass[twocolumn,showpacs,preprintnumbers,amsmath,amssymb]{revtex4}

\usepackage{graphicx}
\usepackage{bm}

\usepackage{color}

\begin{document}

\title{Note and calculations concerning elastic dilatancy in 2D glass-glass liquid foams}

\author{Fran\c{c}ois Molino}
\author{Pierre Rognon}
\author{Cyprien Gay}

\email{cyprien.gay@univ-paris-diderot.fr}

\affiliation{%
Centre de Recherche Paul Pascal, CNRS, UPR~8641,
Universit\'e de Bordeaux~1,
115 Av. Schweitzer, F--33600 PESSAC, France\\
Department of Endocrinology,
Institute of Functional Genomics, CNRS, UMR~5203,
INSERM U661, Universities of Montpellier~1 and~2,
141 rue de la Cardonille, F--34094 MONTPELLIER cedex 05, France\\
Mati\`{e}re et Syst\`{e}mes Complexes, 
Universit\'{e} Paris Diderot--Paris 7, CNRS, UMR~7057, 
B\^atiment Condorcet, Case courrier 7056, F--75205 PARIS cedex 13, France
}%

\date{\today}

\begin{abstract}
When deformed, liquid foams tend to raise their liquid contents
like immersed granular materials, a phenomenon called dilatancy.
We have aready described a geometrical interpretation
of elastic dilatancy in 3D foams 
and in very dry foams squeezed between two solid plates (2D GG foams).
Here, we complement this work
in the regime of less dry 2D GG foams.
In particular, we highlight the relatively strong dilatancy effects
expected in the regime where we have predicted
rapid Plateau border variations.
\end{abstract}

\pacs{47.20.Dr, 83.80.Iz, 47.57.Bc, 68.03.Cd}
\keywords{foam - emulsion - materials - Plateau border - volume fraction}
\maketitle

\newcommand{\hide}[1]{}

\newcommand{\hs}{\hspace{0.6cm}}
\newcommand{\be}{\begin{equation}}
\newcommand{\ee}{\end{equation}}
\newcommand{\bee}{\begin{eqnarray}}
\newcommand{\eee}{\end{eqnarray}}

\newcommand{\transp}[1]{{#1}^T}
\newcommand{\trace}{{\rm tr}}

\newcommand{\si}{\sigma}
\newcommand{\unittensor}{I}
\newcommand{\surfacetension}{\gamma}
\newcommand{\specificsurface}{\Sigma}
\newcommand{\siinterf}{\si^{\rm interf}}
\newcommand{\philiq}{\phi}
\newcommand{\deform}{\epsilon}
\newcommand{\perim}{L}

\newcommand{\dimratio}{k}
\newcommand{\truebidi}{E-F}

\newcommand{\pgas}{p_{\rm g}}
\newcommand{\pliq}{p_{\rm l}}
\newcommand{\Dp}{\Delta p}
\newcommand{\pwall}{p_{\rm wall}}
\newcommand{\pdis}{\Pi_{\rm d}}
\newcommand{\piosm}{\pi_{\rm osm}}
\newcommand{\dilatancy}{\chi}

\newcommand{\Sbub}{S} 
\newcommand{\Abub}{{\cal A}} 
\newcommand{\Atot}{{\cal A}_{\rm tot}} 
\newcommand{\Vbub}{\Omega} 
\newcommand{\Vliq}{\Omega_{\rm liq}} 
\newcommand{\Vpb}{V_{\rm Pb}} 
\newcommand{\sppb}{S_{\rm hPb}} 
\newcommand{\Pbradius}{R} 
\newcommand{\pPbradius}{\Pbradius_{\rm ps}}  

\newcommand{\Vwn}{V_{\rm wn}} 
\newcommand{\shpb}{\sppb} 
\newcommand{\svpb}{S_{\rm vPb}} 

\newcommand{\va}{\vec{a}}
\newcommand{\vb}{\vec{b}}
\newcommand{\vc}{\vec{c}}
\newcommand{\vu}{\vec{u}}
\newcommand{\vv}{\vec{v}}
\newcommand{\vw}{\vec{w}}

\section{Dilatancy in foams}\label{Sec:Intro}

\begin{figure}
\begin{center}
\resizebox{1.0\columnwidth}{!}{%
\includegraphics{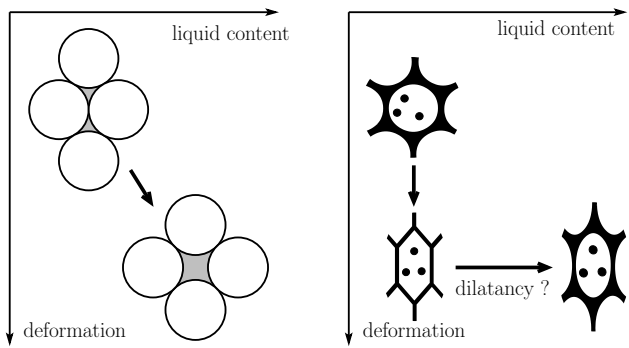}
}
\end{center}
\caption{Dilatancy in granular media and in foams.
When a granular material (left), 
initially in a compact state, is deformed, 
the steric interactions between grains
cause the liquid content to rise.
By contrast, in a foam (right),
because the bubbles are deformable,
the liquid volume fraction may remain constant.
Static dilatancy means that if the liquid fraction
tends to increase due to the applied deformation.
}
\label{Fig:grains_foam_deformation_versus_liquid_content}
\end{figure}

Liquid foams~\cite{weaire_hutzler_1999_book,french_book_belin_2010} 
and granular materials both exhibit  ``dilatancy'',
described by Reynolds~\cite{reynolds_centenary_1985_469}
in the context of granular materials:
upon deformation, because grains are forced
to move while avoiding each other,
the medium swells to some extent. 
In other words, the fluid volume fraction $\philiq$ is increased
(see Fig.~\ref{Fig:grains_foam_deformation_versus_liquid_content}).
This effect can remain unnoticed in air.
By contrast, a spectacular absorption of liquid~\cite{bagnold_1941}
is obtained upon deformation of an immersed granular sample.

Because bubbles can deform individually,
a foam might deform substantially
without altering its liquid fraction
(see Fig.~\ref{Fig:grains_foam_deformation_versus_liquid_content}).
Hence, there is {\em a priori} no reason
why a foam should display dilatancy.
In fact, dilatancy {\em does exist} in foams.
This has been shown both experimentally~\cite{marze_2005_121}
in the case of ``dynamic'' dilatancy
(caused by a continuous foam deformation)
and theoretically~\cite{weaire_2003_2747,rioual_2005_117,dilatancy_geometry_letter_2008} 
in the case of ``elastic'' dilatancy
(caused by a constant deformation of the foam).

Dilatancy can be defined~\cite{weaire_2003_2747}
in terms of the osmotic pressure 
of a foam confined in a container,
which corresponds to the force that must be applied externally
to one of the confining walls if the latter
is permeable to the liquid but not to the bubbles.
The static dilatancy coefficient $\dilatancy$
reflects the fact that the foam osmotic pressure
varies with the deformation $\epsilon$
when the fluid volume fraction $\philiq$ is kept constant:
\be
\label{Eq:def_dilatancy_derivative}
\dilatancy=\left.\frac{\partial^2\piosm}{\partial\epsilon^2}\right|_\philiq
\ee
Surprisingly, this coefficient can be either {\em positive}
like in granular materials or {\em negative}~\cite{weaire_2003_2747}.
In the case of a negative dilatancy coefficient,
deforming the foam results in a tendency
to expell liquid and make the foam dryer.

We have shown that the origin of this change in sign
can be traced back to two different 
physical contributions~\cite{dilatancy_geometry_letter_2008}.
On the one hand, deforming the foam
implies an increase in the total of all
Plateau border lengths in the sample
(in the case of a 2D GG foam~\cite{Vaz_Cox_2005_415}, 
these are the {\em pseudo} Plateau borders,
{\em i.e.}, those along the solid plates).
As a result of this increase in length,
when the total amount of liquid is kept constant,
the Plateau borders shrink,
which raises the pressure difference
between the gas and the liquid,
resulting in an increase of the osmotic pressure,
hence a positive contribution to dilatancy.
On the other hand, deforming the foam
also implies an increase in the total surface area
of the gas-liquid interfaces.
Because the interfaces contribute negatively
to the stress tensor in the foam (tensile contribution),
this has a negative contribution 
to the osmotic pressure, and hence, to dilatancy.
Because the increase in total interfacial energy
is directly related to the elasticity of the foam,
this negative contribution to dilatancy coefficient 
is proportional to the elastic modulus,
as shown by Weaire and Hutzler~\cite{weaire_2003_2747},

In the present work, 
with this geometrical interpretation in mind,
we conduct the explicit calculation
of the elastic dilatancy 
of not too dry 2D GG foams (regimes A, B, C and D).
In particular, we highlight the relatively strong dilatancy effects
expected in the regime where we have predicted
rapid Plateau border variations.

\section{Geometry of 2D GG foams: floor tile versus pancake regime}\label{Sec:geometry}

Let us first choose notations~\cite{dilatancy_geometry_letter_2008,bulimia_2010}
to describe the geometry 
of two-dimensional foams squeezed between two glass plates,
called ``GG'' foams.

\begin{figure}
\begin{center}
\resizebox{1.0\columnwidth}{!}{%
\includegraphics{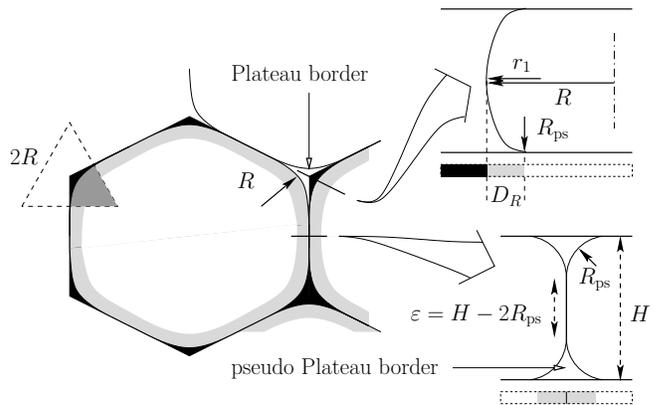}
}
\end{center}
\caption{Pancake conformation of a bubble
squeezed between two solid plates (distance $H$): 
approximate geometry.
Left: top view.
The variable $L$ denotes 
the average perimeter of the bubbles in such a top view
(outer perimeter of the light grey region).
The variable $\Vbub$ is the volume of the bubble gas
(full thickness of the white region
and part of the thickness of the light grey region),
and $\Abub$ is defined as $\Abub=\Vbub/H$.
The variable $\Atot$ is the total (gas and liquid)
projected surface area per bubble (white and light grey and black regions).
The Plateau rules imply that the angle
of the medium grey sector is $\pi/3$.
Right: two different cross-sections (side views)
with matching greyscale.
As seen from above, the cont
act between two bubbles
is typically along a straight line (left).
The pseudo Plateau borders then have a uniform curvature
in this region (radius $\pPbradius$, see bottom right drawing).
By contrast, in the Plateau border region, 
the section of the gas-liquid interface
is approximately elliptical in shape (top right drawing), 
with radii of curvature $r_1$ at mid-height in the vertical direction
and $\pPbradius$ at the plates,
while the radius of curvature at mid-height 
in the horizontal direction is $\Pbradius$.
The width $D_R$ of the curved region
is intermediate between $\pPbradius$ and $H/2$
while $r_1$ is larger than $H/2$.
}
\label{Fig:pancake}
\end{figure}

We call $L$ the perimeter of the bubble,
defined as the perimeter of the rounded polygon
that constitutes the bubble, as seen from above,
and $\Pbradius$ the corresponding radius of curvature
of the Plateau border (see Fig.~\ref{Fig:pancake}, left).
We call $H$ the distance between both solid plates,
and $\pPbradius$ the radius of curvature 
of the pseudo Plateau borders 
(see Fig.~\ref{Fig:pancake}, bottom right).
We also call $\Vbub$ the average bubble volume,
and $\Vliq$ the average volume of liquid per bubble.
The liquid volume fraction $\philiq$ thus verifies:
\be
\label{Eq:Omega_liq}
\philiq=\frac{\Vliq}{\Vbub+\Vliq},
\hs{i.e.}\hs
\Vliq=\Vbub\frac{\philiq}{1-\philiq}.
\ee
We assume that $\Vbub$ remains constant:
the applied stresses are not sufficient
to compress the gas phase significantly
unless the bubble size is on the order of a micron.

\begin{figure}
\begin{center}
\resizebox{0.8\columnwidth}{!}{%
\includegraphics{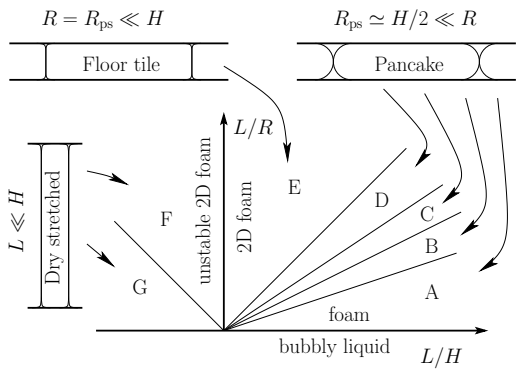}
}
\end{center}
\caption{Regimes of a glass-glass 2D foam 
with low liquid fraction foam ($\philiq\ll 1$),
in terms of the bubble perimeter $L$,
the Plateau border radius $\Pbradius$
and the cell height $H$.
The bubble perimeter $L$ is measured at mid-height of the cell:
it is the outer perimeter of the light grey ribbon
in Fig.~\ref{Fig:pancake}.
Such a foam can be found in two main configurations.
In the floor tile situation (regimes E-G)
the pseudo Plateau borders
are much thinner than the cell height
($\Pbradius=\pPbradius\ll H$).
By contrast, in the pancake regime,
although the overall liquid volume fraction $\philiq$
is still much smaller than 1,
facing pseudo Plateau borders almost join
($H-2\pPbradius\ll H\ll\Pbradius$),
and each bubble has a pancake-like shape.
More precisely, when considering
properties such as dilatancy,
it is useful to subdivide the pancake situation
into regimes A-D defined
by Table~\ref{tab:geometrical_transitions}.
The corresponding expressions 
for the liquid volume fraction and for dilatancy 
are indicated on Tables~\ref{tab:geom_gg}
and~\ref{tab:results}.
Note that regimes F and G,
where the height is larger than the perimeter,
do not always correspond to stable 2D GG foams,
as shown in Ref.~\cite{cox_epje2002_311}.
The limit of a truly two-dimensional foam
is obtained in regime G with $H\rightarrow\infty$.}
\label{Fig:diagram_pancake_regimes_ABCD}
\end{figure}

\begin{table}
\begin{tabular}{ | c | c | c | c | c | }
\hline
$\philiq<\philiq_c$
& $\Pbradius<L$
& A
& $\Pbradius^3 > L^2\,H$
& -
\\
\hline
-
& $\Pbradius^3 < L^2\,H$
& B
& $\Pbradius^2 > L\,H$
& $\Omega_{\rm Pb}>\Omega_{\rm pPb}$
\\
\hline
$\Omega_{\rm Pb}<\Omega_{\rm pPb}$
& $\Pbradius^2 < L\,H$
& C
& $\Pbradius^3 > L\,H^2$
& -
\\
\hline
-
& $\Pbradius^3 < L\,H^2$
& D
& $\Pbradius > H$
& pancake
\\
\hline
floor tile
& $\Pbradius < H$
& E
& $L > H$
& 2D
\\
\hline
stretched 2D
& $L < H$
& F
& $L^2 > H\,\Pbradius$
& -
\\
\hline
-
& $L^2 < H\,\Pbradius$
& G
& $L > \Pbradius$
& $\philiq<\philiq_c$
\\
\hline
\end{tabular} 
\caption{Seven regimes for a 2D glass-glass foam.
Regimes A-D correspond to pancake-shaped bubbles,
while regime E corresponds to a foam made of floor tile shaped bubbles.
In regimes C-E, most of the liquid
is located in the pseudo Plateau borders,
whereas in regime A and B, 
the Plateau borders themselves have a greater volume.
The transitions between regimes A and B,
and that between C and D 
(which are meaningful as far as dilatancy is concerned)
have no simple geometrical interpretation.
Note that regimes F and G,
where the height is larger than the perimeter,
do not always correspond to stable 2D GG foams,
as shown in Ref.~\cite{cox_epje2002_311}.
But the limit $H\rightarrow\infty$ in regime G
corresponds to an ideal 2D foam.}
\label{tab:geometrical_transitions}
\end{table}

The main contributions to the quantity of liquid per bubble
are pictured on Fig.~\ref{Fig:pancake}
and can be calculated from simple geometrical arguments:
\be
\label{Eq:Vliq_PR}
\Vliq \simeq
(2-\pi/2)\,L\,\pPbradius^2
+(2\sqrt{3}-\pi)\,\Pbradius^2 H
\ee
The first term corresponds 
to the pseudo Plateau borders,
which make the junction between the interbubble films
and the solid plates.
As seen from above,
they correspond to the light grey regions
in Fig.~\ref{Fig:pancake}.
Each portion of their interfaces
has the shape of a quarter of a circular cylinder
(see Fig.~\ref{Fig:pancake}, bottom right).
The second term in Eq.~(\ref{Eq:Vliq_PR})
corresponds to the genuine Plateau borders
(black regions in Fig.~\ref{Fig:pancake}),
whose three contours span an angle $\pi/3$ each.

Eq.~(\ref{Eq:Vliq_PR}) indicates
that the squeezed 2D foam can be found in two main regimes
depending on volume fraction and geometry.
They are pictured on Fig.~\ref{Fig:diagram_pancake_regimes_ABCD}.

\begin{enumerate}
\item When the Plateau border radius $\Pbradius$ is much larger
than the sample thickness $H$ 
(regimes A-D of Fig.~\ref{Fig:diagram_pancake_regimes_ABCD}),
each bubble takes the form of a thick ``pancake'',
and its edge is like a half cylinder with radius $H/2$.
\item In the reverse limit, the bubbles
are shaped more like ``floor tile'', with sharp edges
(regime E of Fig.~\ref{Fig:diagram_pancake_regimes_ABCD}):
this time, the Plateau borders are like fine threads
pinned on both solid plates,
and each pseudo Plateau borders resembles
a stretched, fine thread, glued on one of the solid plate
and joining the attachment points of two Plateau borders.
\end{enumerate}
In Fig.~\ref{Fig:diagram_pancake_regimes_ABCD},
we have also pictured regimes F and G:
they are useful to obtain the limit of ideal 2D foams,
for which the solid plates are so far apart
that the volume of the pseudo Plateau borders
can be entirely neglected.
Regime G is useful in particular for 
dilatancy~\cite{dilatancy_geometry_letter_2008}
as it reproduces the negative dilatancy result
obtained in the very dry limit~\cite{weaire_2003_2747}.
Note that as shown by Cox, Weaire and Vaz
both analytically and numerically~\cite{cox_epje2002_311},
any bubbles with less than six neighbours
will tend to gather on one of the solid plates for large separations,
thus turning the foam into a three-dimensional rather than two-dimensional foam.
As a result, in the case where all bubbles inside the foam
have exactly six or more (hence exactly six) neighbours,
the plates can be separated indefinitely
without triggering any rearrangements.

In the pancake regime (A-D),
one has $\pPbradius\simeq H/2$.
Taking into account the elliptical shape
of the Plateau border cross-section
taken perpendicularly to the plates
(see Ref.~\cite{bulimia_2010} for details),
the volume of liquid per bubble can be expressed as:
\be
\label{Eq:Vliq_PR_ABCD_refined}
\Vliq^{\rm ABCD}\simeq 
\frac{4-\pi}{8}\,H^2L
\,\left[1-\frac{2H}{3\Pbradius}\right]
+(2\sqrt{3}-\pi)\,\Pbradius^2H
\ee

In regimes A and B, the Plateau border contribution dominates:
$\Vliq \simeq (2\sqrt{3}-\pi)\,\Pbradius^2 H$.
Conversely, in regimes C and D,
the pseudo Plateau borders contain most of the liquid:
$\Vliq \simeq (1/2-\pi/8)\,L\,H^2$.

In all three regimes E-G ($\Pbradius\ll H$, floor tile regime), 
the radius of curvature $\pPbradius$
of the pseudo Plateau borders 
is equal to that of the Plateau borders $\pPbradius^{\rm EFG}=\Pbradius$
and Eq.~(\ref{Eq:Vliq_PR}) reduces to:
\be
\label{Eq:Vliq_PR_EFG}
\Vliq^{\rm EFG} \simeq
[(2-\pi/2)\,L+(2\sqrt{3}-\pi)\,H]\,\Pbradius^2
\ee

\section{Dilatancy in the CD regime}

In this Section, we focus on regimes C and D,
see Fig.\ref{Fig:diagram_pancake_regimes_ABCD}
They are defined by the following conditions:
\begin{center}
\begin{tabular}{|c|c|c|}
\hline
$R^2<LH$ & C & $R^3>LH^2$\nonumber \\
\hline
$R^3<LH^2$ & D & $R>H$\\
\hline
\end{tabular}\\
\end{center}
Indeed, we have shown
that in these regimes, the size of the Plateau borders
change in a particularly rapid manner~\cite{bulimia_2010} 
as a function of volume fraction $\philiq$, 
inter-plate distance $H$ or bubble size $L$.
We thus expect stronger dilatancy effects in these two regimes.
We will demonstrate below that it is indeed the case.

In order to conduct the corresponding calculation,
our first aim is to obtain an expression relating
the variation $\delta\philiq$ of the liquid volume fraction, 
the variation $\delta\piosm$ of the osmotic pressure and the variation $\delta L$ of the perimeter (related to the foam deformation $\epsilon$).

We start from 
\be
\phi^{CD}\simeq\frac{4-\pi}{8}\frac{H^2L}{\Omega}
\ee
\bee
\pi^{CD}_{zz}&\simeq& \frac{2\gamma}{H}\nonumber \\
\pi^{CD}_{pl}&\simeq& \frac{2}{3}\frac{\gamma}{R}
\eee
which are obtained (in regimes C and D)
from Eqs.~(\ref{Eq:philiq_ABCD_Vbub})
and~(\ref{Eq:piosm_zz_ABCD}-\ref{Eq:piosm_pl_ABCD}).

Differentiating Eqs.~(\ref{Eq:philiq_ABCD_Vbub}) and~(\ref{Eq:piosm_zz_ABCD}-\ref{Eq:piosm_pl_ABCD}), we obtain:
\be
\delta\, \phi^C\; =\;\frac{4-\pi}{8}\frac{H^2}{\Omega}\delta L\;+\; (4\sqrt{3}-2\pi)\frac{RH}{\Omega}\delta R,
\ee
\be
\delta\, \phi^D\; =\;\frac{4-\pi}{8}\frac{H^2}{\Omega}\delta L\;+\; \frac{4-\pi}{12}\frac{H^3 L}{\Omega R^2}\delta R,
\ee
\bee
\delta \pi_ {zz}\; + \;  \frac{2\gamma}{H} \,\delta\, \phi &=& -\frac{2}{3}\frac{\gamma}{R^2}\,\delta R -\frac{\pi}{4} \, \frac{\gamma H}{\Omega} \delta L,\nonumber \\
\delta \pi_ {pl}\; + \;  \frac{2\gamma}{R} \,\delta\, \phi &=& -\frac{2}{3}\frac{\gamma}{R^2}\,\delta R -\frac{\pi}{8} \, \frac{\gamma H}{\Omega} \delta L.
\eee

Eliminating $\delta R$ between the equations for $\delta\phi$ and $\delta\phi$, we obtain:

\be
\delta \pi^C_ {zz,pl}\; + \; \frac{2}{3(4\sqrt{3}-2\pi)}\; \frac{\gamma \Omega}{R^3 H} \delta \phi \;=\;  \frac{4-\pi}{12(4\sqrt{3}-2\pi)}\; \frac{\gamma H}{R^3} \delta L, 
\ee
\be
\delta \pi^D_ {zz,pl}\;=\;-\frac{8}{4-\pi}\,\frac{\gamma \Omega}{H^3 L} \delta\phi\;+\; \frac{\gamma}{HL}
\delta L.
\ee
As shown in Ref.~\cite{dilatancy_geometry_letter_2008},
the variation of the perimeter
is related to the foam deformation 
in the following way, see Appendix~\ref{App:bubble_perimeter}:
\be
\frac{\delta L}{L} \, = \, \frac{\epsilon ^2}{2}
\ee
Then, using 
\be
\Omega=L^2 H,
\ee
\be
\frac{LH^2}{\Omega}=\frac{H}{L}\ll1,
\ee
and
\be
\frac{LHR}{\Omega}=\frac{R}{L}\ll 1,
\ee
we get:
\be
\delta \pi^C_ {zz,pl}\; + \; \frac{2}{3(4\sqrt{3}-2\pi)}\; \frac{\gamma \Omega}{R^3 H} \delta \phi \;=\;  \frac{4-\pi}{24(4\sqrt{3}-2\pi)}\; \frac{\gamma H L}{R^3} \epsilon ^2
\ee
\be
\delta \pi^D_ {zz,pl}\;+\;\frac{8}{4-\pi}\,\frac{\gamma \Omega}{H^3 L} \delta\phi\;=\; \frac{1}{2}\frac{\gamma}{H}\epsilon ^2
\ee
Setting $\delta\phi=0$, we obtain
the dilatancy coefficients defined
by Eq.~(\ref{Eq:def_dilatancy_derivative}),
which characterize the immediate change
in osmotic pressure due to the foam deformation:
\be
\chi^C\,=\,\frac{4-\pi}{12(4\sqrt{3}-2\pi)} \frac{\gamma H L}{R^3},
\ee
\be
\chi^D\,=\,\frac{\gamma}{H}.
\ee
Conversely, setting $\delta\pi=0$,
we obtain the eventual change in liquid fraction
that results from a deformation $\epsilon$:
\be
\delta \phi^{CD}\,=\,\frac{4-\pi}{16}\frac{H^2 L}{\Omega} \epsilon ^2.
\ee
Using
\be
\phi ^C\;=\; \phi^D\;=\; \frac{4-\pi}{8}\frac{H^2 L}{\Omega},
\ee
we get the relative change in liquide volume fraction:
\be
\frac{\delta \phi}{\phi}\;=\;\frac{1}{2}\epsilon ^2 \;=\;\frac{\delta L}{L}.
\ee

\begin{table}
\begin{tabular}{ | c c | c | c | }
\hline
Quantity & Eqs. & Value & Regimes \\
\hline
pseudo Plateau & $\pPbradius$
 & $\frac{H}{2}\,\left(1-\frac{H}{3\Pbradius}\right)$ & ABCD \\
border radius & Ref.~\cite{bulimia_2010}
 & $\Pbradius$ & EFG \\
\hline
volume & & $(2\sqrt{3}-\pi)\,\Pbradius^2\,H$ & ABFG \\
of liquid & $\Vliq$
 & $\frac{4-\pi}{8}\,L\,H^2$ & CD \\
per bubble & (\ref{Eq:Vliq_PR_ABCD_refined},\,\ref{Eq:Vliq_PR_EFG})
 & $\frac{4-\pi}{2}\,L\,\Pbradius^2$ & E \\
\hline
liquid & & $(2\sqrt{3}-\pi)\,\frac{\Pbradius^2}{\Atot}$ & ABFG \\
volume fraction & $\philiq$ 
& $\frac{4-\pi}{8}\,\frac{L\,H}{\Atot}$ & CD \\
($\philiq=\frac{\Vliq}{\Atot\,H}$) 
& (\ref{Eq:philiq_ABCD}-\ref{Eq:philiq_EFG_Vbub})
 & $\frac{4-\pi}{2}\,\frac{\Pbradius^2\,L}{\Atot\,H}$ & E \\
\hline
& & $\frac{2}{H}-\frac{(4\sqrt{3}-2\pi)\,\Pbradius^2}{\Atot\,H}$ & AB \\
specific & $\specificsurface$ & $\frac{2}{H}+\frac{\pi-2}{2}\,\frac{L}{\Atot}$ & CD \\
surface area & Ref.~\cite{bulimia_2010}
& $\frac{2}{H}+\frac{L}{\Atot}$ & E \\
& & $\frac{L}{\Atot}$ & FG \\
\hline
normal & $\piosm^{zz}$ 
 & $2\,\frac{\surfacetension}{H}
+\frac23\,\frac{\surfacetension}{\Pbradius}
-\frac{\pi}{4}\,\frac{\surfacetension\,L\,H}{\Vbub}$ & ABCD \\
osmotic pressure & (\ref{Eq:piosm_zz_ABCD},\,\ref{Eq:piosm_zz_EFG})
& $\frac{\surfacetension}{\Pbradius}
-\frac{\surfacetension\,L\,H}{\Vbub}$ & EFG \\
\hline
in-plane & $\piosm^{pl}$
 & $\frac23\,\frac{\surfacetension}{\Pbradius}
-\frac{\pi}{8}\,\frac{\surfacetension\,L\,H}{\Vbub}$ & ABCD \\
osmotic pressure & (\ref{Eq:piosm_pl_ABCD},\,\ref{Eq:piosm_pl_EFG})
& $\frac{\surfacetension}{\Pbradius}
-2\,\frac{\surfacetension}{H}
-\frac{1}{2}\,\frac{\surfacetension\,L\,H}{\Vbub}$ & EFG \\
\hline
shear & $G$
 & $\frac{\pi}{16}\,\frac{\surfacetension\,L\,H}{\Vbub}$ & ABCD \\
modulus & (\ref{Eq:shear_modulus_ABCD}-\ref{Eq:shear_modulus_EFG})
 & $\frac{1}{4}\,\frac{\surfacetension\,L\,H}{\Vbub}$ & EFG \\
\hline
\end{tabular} 
\caption{Geometrical and stress properties
of a two-dimensional glass-glass foam.
The numbers refer to the relevant series of equations
and the letters to the regimes
of Fig.~\ref{Fig:diagram_pancake_regimes_ABCD}:
pancake regime (A-D) and floor tile regime (E-G).
}
\label{tab:geom_gg}
\end{table}

\begin{table}
\begin{tabular}{ | c c | c | c | }
\hline
Quantity & Eqs. & Value & Regimes \\
\hline
Plateau & & $-\frac{4-\pi}{32\sqrt{3}-16\pi}\,\frac{LH}{\Pbradius^2}$ & ABC \\
border & $\left.\frac{\delta\Pbradius/\Pbradius}{\delta L/L}\right|_\philiq$
 & $-\frac32\,\frac{\Pbradius}{H}$ & D \\
variation & (\ref{Eq:delta_R_fonction_delta_P_ABCD},\,\ref{Eq:delta_R_fonction_delta_P_EFG})
 & $-\frac12$ & E \\
& & $-\frac{4-\pi}{8\sqrt{3}-4\pi}\,\frac{L}{H}$ & FG \\
\hline
& & $-\frac{\pi}{4}\,\frac{\surfacetension\,L\,H}{\Vbub}=-4\,G^{\rm ABCD}$ & A \\
Normal & 
& $\frac{4-\pi}{12(4\sqrt{3}-2\pi)}\,\frac{\surfacetension\,H\,L}{\Pbradius^3}$ & BC \\
elongational & $\left.\frac{\delta\piosm^{zz}}{\delta L/L}\right|_\philiq$ 
& $\frac{\surfacetension}{H}$ & D \\
dilatancy & (\ref{Eq:delta_piosm_zz_ABCD},\,\ref{Eq:delta_piosm_zz_EFG})
& $\frac12\,\frac{\surfacetension}{\Pbradius}$ & E \\
$\dilatancy^{\rm el}_{zz}$ & & $\frac{4-\pi}{8\sqrt{3}-4\pi}\,\frac{\surfacetension\,L}{\Pbradius\,H}$ & F \\
& & $-\frac{\surfacetension\,H\,P}{\Vbub}=-4\,G^{\rm EFG}$ & G \\
\hline
& & $-\frac{\pi}{8}\,\frac{\surfacetension\,L\,H}{\Vbub}=-2\,G^{\rm ABCD}$ & A \\
In-plane & 
& $\frac{4-\pi}{12(4\sqrt{3}-2\pi)}\,\frac{\surfacetension\,H\,L}{\Pbradius^3}$ & BC \\
elongational & $\left.\frac{\delta\piosm^{pl}}{\delta L/L}\right|_\philiq$ 
& $\frac{\surfacetension}{H}$ & D \\
dilatancy & (\ref{Eq:delta_piosm_pl_ABCD},\,\ref{Eq:delta_piosm_pl_EFG})
& $\frac12\,\frac{\surfacetension}{\Pbradius}$ & E \\
$\dilatancy^{\rm el}_{pl}$& & $\frac{4-\pi}{8\sqrt{3}-4\pi}\,\frac{\surfacetension\,L}{\Pbradius\,H}$ & F \\
& & $-\frac12\,\frac{\surfacetension\,H\,L}{\Vbub}=-2\,G^{\rm EFG}$ & G \\
\hline
Liq. fraction & & $-\frac{\pi}{4(4\sqrt{3}-2\pi)}\,\frac{L\,H}{\Pbradius^2}$ & A \\
variation & & $\frac{4-\pi}{4(4\sqrt{3}-2\pi)}\,\frac{L\,H}{\Pbradius^2}$ & B \\
at constant & $\left.\frac{\delta\philiq/\philiq}{\delta L/L}\right|_{\piosm^{zz}}$ & $1$ & CDE \\
normal & (\ref{Eq:delta_philiq_over_philiq_over_delta_P_over_P_ABCD_zz},\,
\ref{Eq:delta_philiq_over_philiq_over_delta_P_over_P_EFG_zz})
 & $\frac{4-\pi}{4\sqrt{3}-2\pi}\,\frac{L}{H}$ & F \\
$\piosm^{zz}$ & & $-2\,\frac{L\,H\,\Pbradius}{\Vbub}$ & G \\
\hline
Liq. fraction & & $-\frac{3\pi}{8}\,\frac{L\,H\,\Pbradius}{\Vbub}$ & A \\
variation & & $\frac{4-\pi}{4(4\sqrt{3}-2\pi)}\,\frac{L\,H}{\Pbradius^2}$ & B \\
at constant & $\left.\frac{\delta\philiq/\philiq}{\delta L/L}\right|_{\piosm^{pl}}$
 & $1$ & CDE \\
in-plane & (\ref{Eq:delta_philiq_over_philiq_over_delta_P_over_P_ABCD_pl},\,
\ref{Eq:delta_philiq_over_philiq_over_delta_P_over_P_EFG_pl})
 & $\frac{4-\pi}{4\sqrt{3}-2\pi}\,\frac{L}{H}$ & F \\
$\piosm^{pl}$ & & $-\frac{L\,H\,\Pbradius}{\Vbub}$ & G \\
\hline
\end{tabular} 
\caption{Predictions for dilatancy. 
The variation of the Plateau border radius
is taken at constant volume fraction.
The elongational dilatancy coefficient
is related to the shear coefficient
through Eq.~(\ref{Eq:dilatancy_shear}).
For both the variation of the liquid fraction
at constant osmotic pressure 
and the elongational dilatancy coefficient,
the normal mode as well as the in-plane mode are provided.
The numbers refer to the relevant series of equations
and the letters to the regimes
of Fig.~\ref{Fig:diagram_pancake_regimes_ABCD}.
}
\label{tab:results}
\end{table}

\section{Conclusion}

In the present  follow-up to References~\cite{dilatancy_geometry_letter_2008,bulimia_2010}, 
we derived detailed mechanical properties
of two-dimensional foams squeezed
between parallel solid surfaces.

The main geometrical and mechanical properties of such foams
are summarized in Tables~\ref{tab:geom_gg} and~\ref{tab:results}
and in the corresponding equations indicated therein.

After recalling our geometrical interpretation 
of elastic dilatancy in very dry 2D GG and 3D 
foams~\cite{dilatancy_geometry_letter_2008},
we derived the expected magnitude of such effects
for less dry 2D GG foams.

It must be recalled, at this stage,
that the calculations were carried out
in some asymptotic limits
defined by Table~\ref{tab:geometrical_transitions}.
Some of these regimes (particularly A-D) are rather narrow:
for comparison with a real situation,
it will be more advisable
to use the full equations cited in Table~\ref{tab:results}
than each asymptotic expression listed within the table.
To get more accurate results, 
in particular in regimes C and D,
Surface Evolver simulations would be required
as in Refs.~\cite{Cox_Janiaud_2008_693,bulimia_2010}.

Concerning dilatancy,
the main result of our calculations
is that the expected effect is positive
in most regimes (B-F):
deforming the foam
will induce an {\em increase}
in the osmotic pressure
if the volume is kept constant,
or conversely it will induce
an {\em increase} in the liquid volume fraction
if the osmotic pressure is kept constant.
Only regimes A and G display negative dilatancy~\cite{weaire_2003_2747}:
in these regimes, among both contributions
to the osmotic pressure 
discussed in Ref.~\cite{dilatancy_geometry_letter_2008}
and recalled in the Introduction,
the effect of the increase in total surface area
dominates over the effect of the increase
in total pseudo Plateau border length.

As we mentioned in the Introduction,
the Plateau borders vary rapidly in size
in regimes C and D~\cite{bulimia_2010}.
Let us now discuss how that affects dilatancy.

As can be seen in Table~\ref{tab:results},
the relative change in volume fraction
at constant in-plane osmotic pressure $\piosm^{pl}$,
divided by the square of the deformation, $\epsilon^2=2\delta L/L$
(see Appendix~\ref{App:bubble_perimeter}),
is of order unity in regimes $C$, $D$ and $E$.
It can be checked, using Table~\ref{tab:geometrical_transitions},
that the same quantity is much smaller
in neighbouring asymptotic regimes $B$ and $F$.
The same observation holds at constant
normal osmotic pressure $\piosm^{zz}$.

Regarding both the in-plane and the normal 
dilatancy coefficients $\chi^{\rm el}_{pl}$ and $\chi^{\rm el}_{zz}$,
the same is true.
Indeed, taking their expressions in regime $E$
as a reference, $\chi^{\rm el}(E)\simeq\gamma/R$,
one can check, again using Table~\ref{tab:geometrical_transitions},
that they become bigger in regimes $C$ and $D$
and smaller in $B$ and $F$.

We believe that regimes $S$ and $D$ 
should therefore {\em a priori} constitute
the more promising target for experimental investigations.

\section*{Acknowledgements}

We gratefully acknowledge fruitful discussions
with Benjamin Dollet
and with participants of the GDR 2352 Mousses (CNRS)
and of the Informal Workshop on Foam mechanics (Grenoble 2008).
P.R. was supported by the Agence Nationale de la Recherche (ANR05).

\appendix

\section{Liquid volume fraction}
\label{Sec:liquid_volume_fraction}

From Eqs.~(\ref{Eq:Vliq_PR_ABCD_refined})
and~(\ref{Eq:Vliq_PR_EFG}),
we derive the liquid volume fraction 
$\philiq=\Vliq/(\Atot\,H)$ in the foam,
both in the pancake regime and in the floor tile regime:
\bee
\label{Eq:philiq_ABCD}
\philiq^{\rm ABCD}&\simeq& 
\frac{4-\pi}{8}\,\frac{H\,L}{\Atot}
\,\left[1-\frac{2H}{3\Pbradius}\right]
\nonumber\\
&&+(2\sqrt{3}-\pi)\,\frac{\Pbradius^2}{\Atot}\\
\label{Eq:philiq_EFG}
\philiq^{\rm EFG}&\simeq&
\frac{4-\pi}{2}\,\frac{\Pbradius^2\,L}{\Atot\,H}
+(2\sqrt{3}-\pi)\,\frac{\Pbradius^2}{\Atot}
\eee
The corresponding values of $\philiq$ in all sub-regimes
are indicated in Table~\ref{tab:geom_gg}.
Because the total volume of the bubble and liquid, 
$\Atot\,H=\Vbub+\Vliq$, is not constant, 
it is useful to express the liquid fraction
in terms of the volume $\Vbub$ of the bubble itself.
Using $\philiq/(1-\philiq)=\Atot\,H\,\philiq/\Vbub$,
the above equations become:
\bee
\label{Eq:philiq_ABCD_Vbub}
\frac{\philiq^{\rm ABCD}}{1-\philiq^{\rm ABCD}}&\simeq&
\frac{4-\pi}{8}\,\frac{H^2\,L}{\Vbub}
\,\left[1-\frac{2H}{3\Pbradius}\right]\nonumber\\
&&+(2\sqrt{3}-\pi)\,\frac{\Pbradius^2\,H}{\Vbub}\\
\label{Eq:philiq_EFG_Vbub}
\frac{\philiq^{\rm EFG}}{1-\philiq^{\rm EFG}}&\simeq& 
\frac{4-\pi}{2}\,\frac{\Pbradius^2\,L}{\Vbub}
+(2\sqrt{3}-\pi)\,\frac{\Pbradius^2\,H}{\Vbub}
\eee
%

\section{Bubble perimeter in a crystalline 2D foam}
\label{App:bubble_perimeter}

Let us consider a crystalline foam
subjected to an arbitrary elastic, homogeneous deformation.
Up to a global rotation, it can be expressed as an elongation:
\be
\begin{pmatrix}
\lambda & 0 \\
0 & 1/\lambda
\end{pmatrix}
\ee
The perimeter of such a hexagonal foam 
increases upon deformation in the following 
way~\footnote{Surprisingly,
the increase in bubble perimeter
does not depend on the crystal orientation
with respect to the direction of elongation.} :
\bee
\label{Eq:perimeter_dry}
L^{\rm dry}&=&L^{\rm dry}_0\;
\left[\frac{\lambda}{2}+\frac{1}{2\lambda}\right]\\
\label{Eq:relative_dry_perimeter_increase}
\frac{\delta L}{L^{\rm dry}_0}
&=&\frac{(\lambda-1)^2}{2\lambda}
\eee
\begin{figure}
\begin{center}
\resizebox{1.0\columnwidth}{!}{%
\includegraphics{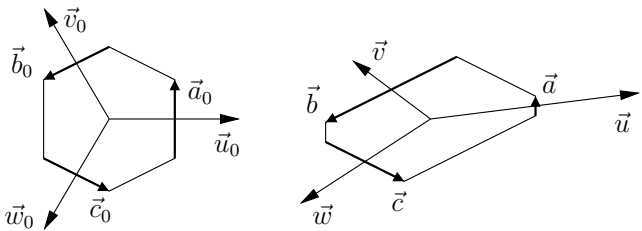}
}
\end{center}
\caption{One hexagonal bubble in a crystalline, 
two-dimensional foam at rest (left) 
and after deformation (right).
Center-to-center vectors $\vu_0$, $\vv_0$ and $\vw_0$
are transported affinely according to the macroscopic deformation,
see Eqs.~(\ref{Eq:uuzero}-\ref{Eq:wwzero}).
By contrast, bubble edges (vectors $\va_0$, $\vb_0$ and $\vc_0$)
rearrange after deformation (vectors $\va$, $\vb$ and $\vc$),
in order to meet at angle $2\pi/3$.
}
\label{Fig:abc_lambda}
\end{figure}

To show it, we consider a crystalline,
two-dimensional foam in the dry limit, 
and derive the total interface contour length
in the foam as a function of the applied deformation.

Let us start with an undeformed foam.
The initial bubble edges $\va_0$, $\vb_0$ and $\vc_0$ 
meet at angle $2\pi/3$ according to Plateau's rule,
and they have identical lengths (hence, $\va_0+\vb_0+\vc_0=\vec{0}$), 
see Fig.~\ref{Fig:abc_lambda}a.
The center-to-center vectors
\bee
\vu_0=\vc_0-\vb_0 \\
\vv_0=\va_0-\vc_0 \\
\vw_0=\vb_0-\va_0,
\eee
which coincide with the principal crystalline axes,
then also meet at angle $2\pi/3$.
When the foam is deformed (transformation $F$),
these vectors are deformed according to:
\bee
\vu &=& F \cdot \vu_0 \label{Eq:uuzero}\\
\vv &=& F \cdot \vv_0 \label{Eq:vvzero}\\
\vw &=& F \cdot \vw_0 \label{Eq:wwzero}
\eee
Correspondingly, the bubble edges
$\va$, $\vb$ and $\vc$ reorganize
so as to not only verify
\bee
\vu=\vc-\vb \label{u_abc}\\
\vv=\va-\vc \label{v_abc} \\
\vw=\vb-\va \label{w_abc},
\eee
but also maintain the $2\pi/3$ angle condition.
This generally implies evolving towards unequal lengths
(see Fig.~\ref{Fig:abc_lambda}b).

For simplicity, we restrict ourselves
to a deformation that conserves the bubble volume
({\em i.e.}, surface area as seen from above):
\bee
\label{Eq:surface_area}
S&=&\frac{\sqrt{3}}{2}(a_0b_0+b_0c_0+c_0a_0)
=\frac{3\sqrt{3}}{2}\;a_0^2 \nonumber\\
&=&\frac{\sqrt{3}}{2}(ab+bc+ca)
\eee
where $a$, $b$ and $c$ are the new edge lengths.
Such a deformation consists in an elongation 
by a factor $\lambda$ in one direction 
and by a factor $1/\lambda$ in the perpendicular direction.
If we fix the direction of vectors $\va_0$ and $\va$
as on Fig.~\ref{Fig:abc_lambda} for convenience,
the most general such transformation can be represented 
by a matrix of the form:
\bee
F=R_2\cdot
\begin{pmatrix}
\lambda&0\\ 0&1/\lambda
\end{pmatrix}
\cdot R_1
\label{Eq:matrixF}
\eee
where $R_1$ and $R_2$ are two rotation matrices.

From Eqs.~(\ref{Eq:uuzero}) to~(\ref{Eq:wwzero})
and~(\ref{Eq:matrixF}),
the center-to-center version~\cite{graner_2008_369}
of the texture tensor~\cite{aubouy_2003_67,asipauskas_2003_71}
can be expressed in matrix form:
\bee
M &=& \frac13[\vu\cdot\transp{\vu}+
\vv\cdot\transp{\vv}+\vw\cdot\transp{\vw}]
\label{M_uvw} \\
&=& \frac13 \; R_2\cdot
\begin{pmatrix}
\lambda&0\\ 0&1/\lambda
\end{pmatrix}
\cdot R_1 \nonumber\\
&&\cdot [\vu_0\cdot\transp{\vu_0}+
\vv_0\cdot\transp{\vv_0}+\vw_0\cdot\transp{\vw_0}] \nonumber\\
&&\cdot R_1^{-1}\cdot
\begin{pmatrix}
\lambda&0\\ 0&1/\lambda
\end{pmatrix}
\cdot R_2^{-1} \\
&=& \frac32 a_0^2\;
R_2 \cdot
\begin{pmatrix}
\lambda^2 &0\\ 0&1/\lambda^2
\end{pmatrix}
\cdot R_2^{-1}
\label{M_azero_lambda}
\eee

From Eq.~(\ref{M_azero_lambda}), we obtain:
\be
\trace(M)= \frac32 a_0^2
\left(\lambda^2+\frac{1}{\lambda^2}\right)
\ee
From Eq.~(\ref{M_uvw})
and Eqs.~(\ref{u_abc}-\ref{w_abc}),
using the fact that vectors
$\va$, $\vb$ and $\vc$
meet at angle $2\pi/3$,
we obtain another expression for $\trace(M)$:
\be
\trace(M)= \frac16 (L^{\rm dry})^2 -3 a_0^2
\ee
where $L^{\rm dry}=2(a+b+c)$ is the bubble perimeter
and where the last term is proportional
to the (fixed) bubble surface area $3\sqrt{3} a_0^2 /2$,
see Eq.~(\ref{Eq:surface_area}).

These two expressions for $\trace(M)$
yield Eq.~(\ref{Eq:perimeter_dry})
for the bubble perimeter $L^{\rm dry}$
in terms of its initial value $L^{\rm dry}_0=6a_0$.

The maximum elongation is $\lambda=\sqrt{3}$ for a few, 
specific orientations of the crystalline network
with respect to the direction of elongation
(it is larger for all other orientations),
and this value
decreases as the foam becomes wetter
as described already long ago~\cite{weaire_hutzler_1999_book}.

\begin{figure}
\begin{center}
\resizebox{1.0\columnwidth}{!}{%
\includegraphics{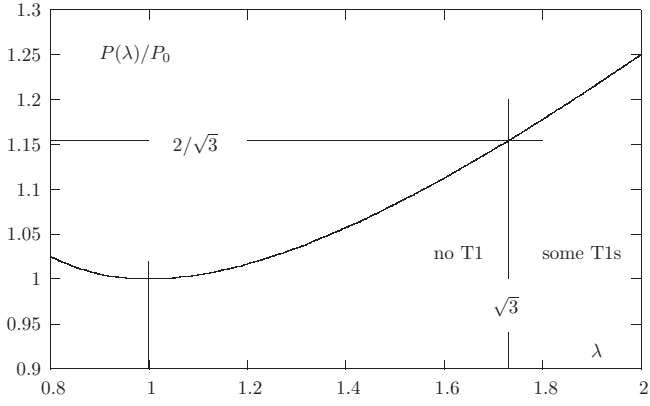}
}
\end{center}
\caption{Bubble perimeter in a crystalline 2D-GG-foam,
as a function of elongation $\lambda$.
The first T1 processes occur when $\lambda=\sqrt{3}$
for specific orientations of the foam.
The corresponding bubble perimeter
is then about $15\%$ longer than at rest.
}
\label{Fig:perimetre}
\end{figure}

\section{Calculation of the asmotic pressure}
\label{Sec:osmotic_pressure_in_2D_GG_foam}

In the present paragraph,
we will calculate the osmotic pressure
of a 2D GG foam.
As mentioned in the Introduction,
it will later be useful
to estimate the foam dilatancy.

When the foam is confined in a container,
the osmotic pressure $\piosm$
corresponds to the force that must be applied externally
to one of the confining walls if the latter
is permeable to the liquid but not to the bubbles.
The osmotic pressure (which is in fact a symmetric tensor
and not just a scalar quantity~\cite{weaire_2003_2747,rioual_2005_117})
is thus the difference between the stress in the foam
and the pressure applied by the pure liquid 
on the other side of the semi-permeable wall:
\be
\label{Eq:piosm_tensor}
\piosm=-\pliq-\si
\ee
(where tensile stresses and pressures are both counted positively).
The stress in the foam 
includes a pressure contribution 
from the liquid ($\pliq$)
and from the gas ($\pgas$),
as well as a tensile contribution from the interfaces:
\be
\si=-\philiq\pliq\unittensor
-(1-\philiq)\pgas\unittensor
+\siinterf. 
\label{Eq:stress_decomposition}
\ee
Hence,
\bee
\piosm&=&(1-\philiq)\,(\pgas-\pliq)-\siinterf\nonumber\\
&=&(1-\philiq)\,\frac{\surfacetension}{\pPbradius}-\siinterf
\label{Eq:piosm_tensor_Dp_siinterf}
\eee

\subsection{Stress due to the interfaces}
\label{Appendix:interface_stress}

Together with the liquid and gas pressures,
the interfaces in a foam contribute 
to the stress in the foam,
as they carry surface tension.
In the present appendix, we derive
a simple expression for this contribution
in the situation of interest.

Each element of interface in the foam,
whose normal is oriented along some vector $\vec{n}$,
has a stress contribution that is in the plane
perpendicular to $\vec{n}$, hence it is proportional to 
$[\unittensor-\vec{n}\otimes\vec{n}]$.

In general terms, let us decompose 
the specific surface area $\specificsurface$ in the foam
(surface area per unit volume)
according to its orientation:
\be
\specificsurface=\iint\specificsurface(\Omega)\;{\rm d}\Omega,
\ee
where ${\rm d}\Omega$ represents an element of solid angle.
The stress contribution from the interfaces
can then be expressed as:
\be
\label{Eq:sinterf}
\siinterf=\surfacetension\iint
[\unittensor-\vec{n}(\Omega)\otimes\vec{n}(\Omega)]\;
\specificsurface(\Omega)\;{\rm d}\Omega
\ee
where $\surfacetension$ is the surface tension.

This implies, in particular, 
that the orientational average
of the interface stress contribution
is simply related to the specific surface:
\bee
<\siinterf>_{3D}&=&\frac13\,\trace(\siinterf)\,\unittensor\nonumber\\
&=&\frac23\,\surfacetension\,\specificsurface\,\unittensor
\eee
as already mentioned through Eq.~(\ref{Eq:siinterf_average_3d}).

We restrict our calculation to the vertical component $\siinterf_{zz}$
and to the in-plane averaged component $\siinterf_{pl}$
of the interfacial stress.

Consider a surface element ${\rm d}S$
whose normal $\vec{n}$ makes an angle $\theta$
with the vertical direction.
If $\theta=0$, the surface element is horizontal
and it contributes $\surfacetension\,{\rm d}S$ towards $\siinterf_{pl}$
and zero towards $\siinterf_{zz}$
By contrast, if $\theta=0$, the surface element is vertical
and it contributes $\frac12\surfacetension\,{\rm d}S$ towards $\siinterf_{pl}$
(where the factor $1/2$ comes from the in-plane orientation average)
and $\surfacetension\,{\rm d}S$ towards $\siinterf_{zz}$.
More generally, it contributes
$\frac{1+\cos^2\theta}{2}\,\surfacetension\,{\rm d}S$ towards $\siinterf_{pl}$
and $\sin^2\theta\,\surfacetension\,{\rm d}S$ towards $\siinterf_{zz}$.

The contribution from the top and the bottom of a bubble
to the in-plane component is:
\bee
\frac{2\,\surfacetension}{H\,\Atot}\,&&\left\{
\Atot-[2\sqrt{3}\,\Pbradius^2-\pi\,(\Pbradius-D_R)^2]
\right.\nonumber\\
&&\left.-(L-2\pi\,\Pbradius)\,\pPbradius
\right\}\nonumber
\eee
Making the approximation $D_R\simeq\pPbradius$
in the Plateau border region~\cite{bulimia_2010} 
and using $\pPbradius\ll L$, this becomes:
\be
\label{Eq:siinterf_pl_both_plates}
\frac{2\,\surfacetension}{H\,\Atot}\,\left\{
\Atot-(2\sqrt{3}-\pi)\,\Pbradius^2
-L\,\pPbradius
\right\}
\ee

The contribution from the vertical films towards $\siinterf_{zz}$ is:
\be
\label{Eq:siinterf_zz_films}
\frac{\surfacetension}{H\,\Atot}\,L\,(H-2\pPbradius)
\ee
Their contribution to $\siinterf_{pl}$ is twice as less
due to the orientation average.

The contribution to $\siinterf_{zz}$ from the menisci,
considered as circular quarter cylinders with radius $\pPbradius$,
can be written as:
\be
\label{Eq:siinterf_zz_menisci}
\frac{\surfacetension}{H\,\Atot}\,
2L\,\int_0^{\frac{\pi}{2}}\sin^2\theta\,\pPbradius\,{\rm d}\theta
=\frac{\surfacetension}{H\,\Atot}\,
\frac{\pi}{2}\,L\,\pPbradius
\ee
Their contribution to $\siinterf_{pl}$
is very similar:
\bee
\frac{\surfacetension}{H\,\Atot}\,
2L\,\int_0^{\frac{\pi}{2}}\frac{1+\cos^2\theta}{2}\,\pPbradius\,{\rm d}\theta
&&\nonumber\\
=\frac{\surfacetension}{H\,\Atot}\,
\frac{3\pi}{4}\,L\,\pPbradius&&
\label{Eq:siinterf_pl_menisci}
\eee

The in-plane interface stress component
thus includes contributions from
Eqs.~(\ref{Eq:siinterf_pl_both_plates}),
(\ref{Eq:siinterf_zz_films})
and (\ref{Eq:siinterf_pl_menisci}):
\bee
\frac{H\,\Atot}{\surfacetension}\,\siinterf_{pl} &\simeq&
2\left[\Atot-(2\sqrt{3}-\pi)\,\Pbradius^2
-L\,\pPbradius\right]\nonumber\\
&&+\frac12\,L\,(H-2\pPbradius)\nonumber\\
&&+\frac{3\pi}{4}\,L\,\pPbradius
\eee
As for the vertical interface stress component,
it includes contributions from
Eqs.~(\ref{Eq:siinterf_zz_films})
and (\ref{Eq:siinterf_zz_menisci}):
\bee
\frac{H\,\Atot}{\surfacetension}\,\siinterf_{zz} &\simeq&
\,L\,(H-2\pPbradius)\nonumber\\
&&+\frac{\pi}{2}\,L\,\pPbradius
\eee

From these two equations, 
we obtain the final results
both in the pancake regime where 
$\pPbradius^{\rm ABCD}\simeq\frac{H}{2}
[1-H/(3R)]$ (see Ref.~\cite{dilatancy_geometry_letter_2008})
and where $\pPbradius\ll\Pbradius$ 
(regimes A-D of Fig.~\ref{Fig:diagram_pancake_regimes_ABCD})
and in the floor tile regime where $\pPbradius=\Pbradius$ (E-G):

\bee
\label{Eq:siinterf_zz_ABCD}
\siinterf_{zz\,\rm ABCD}&\simeq&
\frac{\pi}{4}\,\frac{\surfacetension\,L}{\Atot}
+\frac{4-\pi}{12}\,\frac{\surfacetension\,L\,H}{\Atot\,\Pbradius}\\
\label{Eq:siinterf_pl_ABCD}
\siinterf_{pl\,\rm ABCD}&\simeq&
2\frac{\surfacetension}{H}
-(4\sqrt{3}-2\pi)\frac{\surfacetension\,\Pbradius^2}{\Atot\,H}\nonumber\\
&&+\frac{3\pi-8}{8}\,\frac{\surfacetension\,L}{\Atot}
+\frac{4-\pi}{8}\,\frac{\surfacetension\,L\,H}{\Atot\,\Pbradius}\\
\label{Eq:siinterf_zz_EFG}
\siinterf_{zz\,\rm EFG}&\simeq&
\frac{\surfacetension\,L}{\Atot}
-\frac{4-\pi}{2}\,\frac{\surfacetension\,L\,\Pbradius}{\Atot\,H}\\
\label{Eq:siinterf_pl_EFG}
\siinterf_{pl\,\rm EFG}&\simeq&
2\frac{\surfacetension}{H}
+\frac12\,\frac{\surfacetension\,L}{\Atot}
-\frac{3(4-\pi)}{4}\,\frac{\surfacetension\,L\,\Pbradius}{\Atot\,H}\nonumber\\
&&-(4\sqrt{3}-2\pi)\,\frac{\surfacetension\,\Pbradius^2}{\Atot\,H}
\eee

Note that as expected from the discussion
at the beginning of the present Appendix
and as can be checked from the corresponding expressions 
of the specific surface area~\cite{bulimia_2010},
the average of all three components
is simply related to the total specific surface area $\specificsurface$:
\be
\label{Eq:siinterf_average_3d}
\frac{2\siinterf_{pl}+\siinterf_{zz}}{3}
=\frac23\,\surfacetension\,\specificsurface
\ee
where the numerical factor $2/3$
simply reflects the fact
that each surface element contributes tensile stress
in two out of three directions of space.

\subsection{Expression of the osmotic pressure}

The interfacial contribution $\siinterf$ to the stress
is calculated in Appendix~\ref{Appendix:interface_stress}
with the same geometrical approximations
as the liquid volume given by Eq.~(\ref{Eq:Vliq_PR_ABCD_refined}).
The value is averaged over the sample thickness.
For simplicity, we also average 
in-plane contributions over their orientations.
We thus have one value for the vertical component
and one value for the in-plane averaged component.
The vertical films in the foam contribute
plainly to the vertical component
and partly (due to the orientation average)
to the in-plane component.
The interface in the Plateau and pseudo Plateau borders
make an angle with the vertical direction:
they contribute partly to both components.
As for the top and bottom interfaces of the bubbles,
they contribute plainly to the in-plane components.

The results from Appendix~\ref{Appendix:interface_stress}
are as follows:
\bee
\siinterf_{zz\,\rm ABCD}&\simeq&
\frac{\pi}{4}\,\frac{\surfacetension\,L}{\Atot}
+\frac{4-\pi}{12}\,\frac{\surfacetension\,L\,H}{\Atot\,\Pbradius}\\
%
\siinterf_{pl\,\rm ABCD}&\simeq&
2\frac{\surfacetension}{H}
-(4\sqrt{3}-2\pi)\frac{\surfacetension\,\Pbradius^2}{\Atot\,H}\nonumber\\
&&+\frac{3\pi-8}{8}\,\frac{\surfacetension\,L}{\Atot}
+\frac{4-\pi}{8}\,\frac{\surfacetension\,L\,H}{\Atot\,\Pbradius}\\
%
\siinterf_{zz\,\rm EFG}&\simeq&
\frac{\surfacetension\,L}{\Atot}
-\frac{4-\pi}{2}\,\frac{\surfacetension\,L\,\Pbradius}{\Atot\,H}\\
%
\siinterf_{pl\,\rm EFG}&\simeq&
2\frac{\surfacetension}{H}
+\frac12\,\frac{\surfacetension\,L}{\Atot}
-\frac{3(4-\pi)}{4}\,\frac{\surfacetension\,L\,\Pbradius}{\Atot\,H}\nonumber\\
&&-(4\sqrt{3}-2\pi)\,\frac{\surfacetension\,\Pbradius^2}{\Atot\,H}
\eee

In regimes A-D, 
the first term in Eq.~(\ref{Eq:piosm_tensor_Dp_siinterf})
can be written as:
\be
(1-\philiq)\,\frac{\surfacetension}{\pPbradius}
\simeq (1-\philiq)\,\left[
2\,\frac{\surfacetension}{H}
+\frac23\,\frac{\surfacetension}{\Pbradius}
\right]
\label{Eq:Laplace_pressure_ABCD_un_moins_phi}
\ee
In regimes E-G, it reduces to:
\be
(1-\philiq)\,\frac{\surfacetension}{\pPbradius}
\simeq (1-\philiq)\,\frac{\surfacetension}{\Pbradius}
\label{Eq:Laplace_pressure_EFG_un_moins_phi}
\ee

From Eqs.~(\ref{Eq:philiq_ABCD_Vbub}-\ref{Eq:philiq_EFG_Vbub}),
(\ref{Eq:piosm_tensor_Dp_siinterf}),
(\ref{Eq:siinterf_zz_ABCD}-\ref{Eq:siinterf_pl_EFG})
and~(\ref{Eq:Laplace_pressure_ABCD_un_moins_phi}-\ref{Eq:Laplace_pressure_EFG_un_moins_phi})
and using $\frac{1}{1-\philiq}=\frac{\Atot\,H}{\Vbub}$,
we derive the osmotic pressure:
\bee
\label{Eq:piosm_zz_ABCD}
\frac{\piosm^{zz\,\rm ABCD}}{1-\philiq}&\simeq&
2\,\frac{\surfacetension}{H}+\frac23\,\frac{\surfacetension}{\Pbradius}
-\frac{\pi}{4}\,\frac{\surfacetension\,L\,H}{\Vbub}
\\
\label{Eq:piosm_pl_ABCD}
\frac{\piosm^{pl\,\rm ABCD}}{1-\philiq}&\simeq&
\frac23\,\frac{\surfacetension}{\Pbradius}
-\frac{\pi}{8}\,\frac{\surfacetension\,L\,H}{\Vbub}
\\
\label{Eq:piosm_zz_EFG}
\frac{\piosm^{zz\,\rm EFG}}{1-\philiq}&\simeq&
\frac{\surfacetension}{\Pbradius}
-\frac{\surfacetension\,L\,H}{\Vbub}
+\frac{4-\pi}{2}\,\frac{\surfacetension\,L\,\Pbradius}{\Vbub}\\
\label{Eq:piosm_pl_EFG}
\frac{\piosm^{pl\,\rm EFG}}{1-\philiq}&\simeq&
%
\frac{\surfacetension}{\Pbradius}
-2\,\frac{\surfacetension}{H}
-\frac12\,\frac{\surfacetension\,L\,H}{\Vbub}
\eee

Note that in regime E,
Eq.~(\ref{Eq:piosm_pl_EFG})
yields the expression
$\piosm^{pl\,\rm D}\simeq
(1-\philiq)\,
\left[\frac{\surfacetension}{\Pbradius}
-\frac{\surfacetension\,L\,H}{2\,\Vbub}
\right]
-\frac{2\surfacetension}{H}$
announced earlier~\cite{dilatancy_geometry_letter_2008}.

\section{Calculation of the dilatancy and change in volume fraction}
\subsection{Remark on the calculations}
\label{Sec:simplifications}

In the following paragraphs,
in order to derive such quantities as
the elastic modulus, the dilatancy coefficient
or the change in volume fraction upon deformation,
we will need to differentiate several equations
such as Eqs.~(\ref{Eq:philiq_ABCD_Vbub}-\ref{Eq:philiq_EFG_Vbub})
and~(\ref{Eq:piosm_zz_ABCD}-\ref{Eq:piosm_pl_EFG})
and the expressions for the specific 
surface area~\cite{bulimia_2010}.

When conducting such calculations,
we consider that the gas volume $\Vbub$
as well as the distance $H$ between the solid plates
remain constant.

Once the differentiation is performed,
we determine the relative orders of magnitudes
of the different terms by using such estimates as:
\bee
\Vbub &\propto& L^2\,H\\
\Pbradius &\ll& L\hs({\rm or}\,\,\philiq\ll 1)\\
H &\ll& \Pbradius \hs({\rm ABCD})\\
\Pbradius &\ll& H \hs({\rm EFG})\\
\eee
and simplify the results accordingly.
Performing such simplifications prior to differentiation
would erroneously suppress some relevant terms.

We also assimilate the true perimeter $L$ of the bubble
and the ``dry'' perimeter $L^{\rm dry}$
used to derive Eqs.~(\ref{Eq:perimeter_dry})
and~(\ref{Eq:relative_dry_perimeter_increase}).
The former is the perimeter of the light grey region
in Fig.~\ref{Fig:pancake},
while the latter is the perimeter
that includes the black regions.
Because the Plateau border regions
(medium grey sector in Fig.~\ref{Fig:pancake})
correspond to an angle $\pi/3$,
they are related through:
\be
\label{Eq:definition_perimeter_and_dry_perimeter}
L^{\rm dry}-L=(4\sqrt{3}-2\pi)\,\Pbradius
\ee
Hence, their relative difference
is of order $\Pbradius/L$,
which is small in all regimes A-G
($\philiq\ll 1$).

\subsection{Change in the Plateau border radius}

The foam deformation,
which generates an increase
in the bubble perimeter
given by Eq.~(\ref{Eq:relative_dry_perimeter_increase}),
causes a change in the radius of curvature of the Plateau borders.

From Eq.~(\ref{Eq:philiq_ABCD_Vbub}):
\bee
&&\frac{\delta\philiq^{\rm ABCD}}{(1-\philiq^{\rm ABCD})^2} \simeq
\frac{4-\pi}{8}\,\frac{H^2}{\Vbub}
\,\delta L
\nonumber\\
&&+\left[
(4\sqrt{3}-2\pi)\,\frac{\Pbradius\,H}{\Vbub}
+\frac{4-\pi}{12}\,\frac{H^3\,L}{\Vbub\,\Pbradius^2}
\right]\,\delta\Pbradius
\label{Eq:delta_philiq_ABCD_function_delta_R_and_delta_P}
\eee
Hence, with $\delta\philiq^{\rm ABCD}=0$ and $\philiq^{\rm ABCD}\ll 1$:
\be
\label{Eq:delta_R_fonction_delta_P_ABCD}
\delta\Pbradius^{\rm ABCD}|_{\delta\philiq=0}
\simeq
\frac{-\frac32\frac{\Pbradius^2}{L\,H}}
{1+\frac{12(4\sqrt{3}-2\pi)}{4-\pi}\,\frac{\Pbradius^3}{L\,H^2}}
\,\delta L
\ee

From Eq.~(\ref{Eq:philiq_EFG_Vbub}):
\bee
&&\frac{\delta\philiq^{\rm EFG}}{(1-\philiq^{\rm EFG})^2} \simeq
\frac{4-\pi}{2}\,\frac{\Pbradius^2}{\Vbub}\,\delta L\nonumber\\
&&+\left[(4-\pi)\,\frac{\Pbradius\,L}{\Vbub}
+(4\sqrt{3}-2\pi)\,\frac{\Pbradius\,H}{\Vbub}\right]\,\delta\Pbradius
\label{Eq:delta_philiq_EFG_function_delta_R_and_delta_P}
\eee
Hence, with $\delta\philiq^{\rm EFG}=0$ and $\philiq^{\rm EFG}\ll 1$:
\be
\label{Eq:delta_R_fonction_delta_P_EFG}
\delta\Pbradius^{\rm EFG}|_{\delta\philiq=0}
\simeq\frac{-\frac12\,\frac{\Pbradius}{L}}{1
+\frac{4\sqrt{3}-2\pi}{4-\pi}\,\frac{H}{L}}
\,\delta L
\ee

The results of Eqs.~(\ref{Eq:delta_R_fonction_delta_P_ABCD})
and~(\ref{Eq:delta_R_fonction_delta_P_EFG}) above
are reported in Table~\ref{tab:results}.

\subsection{Shear modulus}

The average bubble surface area,
and hence the foam interfacial energy,
also change as a consequence
of the increase in the bubble perimeter
given by Eq.~(\ref{Eq:relative_dry_perimeter_increase}).
When taken at constant liquid fraction $\philiq$
(or equivalently at constant $\Atot$),
this change is directly related to the shear modulus of the foam.
From the expressions of the specific surface area
obtained in Ref.~\cite{bulimia_2010}, we obtain:

%
\bee
\label{Eq:delta_specific_surface_area_ABCD}
\frac{\left.\delta\specificsurface_{\rm ABCD}\right|_\philiq}
{1-\philiq}&\simeq&
-\left[(8\sqrt{3}-4\pi)\,\frac{\Pbradius}{\Vbub}
+\frac{4-\pi}{6}\,\frac{L\,H^2}{\Vbub\,\Pbradius^2}
\right]\,\delta\Pbradius\nonumber\\
&&+\left[\frac{\pi-2}{2}\,\frac{H}{\Vbub}
+\frac{4-\pi}{6}\,\frac{H^2}{\Vbub\,\Pbradius}
\right]\,\delta L\\
\label{Eq:delta_specific_surface_area_EFG}
\frac{\left.\delta\specificsurface_{\rm EFG}\right|_\philiq}
{1-\philiq}&\simeq&
-\left[(4-\pi)\,\frac{L}{\Vbub}
+(8\sqrt{3}-4\pi)\,\frac{\Pbradius}{\Vbub}
\right]\,\delta\Pbradius\nonumber\\
&&+\left[\frac{H}{\Vbub}
-(4-\pi)\,\frac{\Pbradius}{\Vbub}
\right]\,\delta L
\eee

These equations,
taken at constant volume fraction ($\delta\philiq=0$)
and with $\philiq\ll 1$,
together with Eqs.~(\ref{Eq:delta_R_fonction_delta_P_ABCD})
and~(\ref{Eq:delta_R_fonction_delta_P_EFG}), 
yield:
\bee
\label{Eq:delta_specific_surface_area_ABCD_over_delta_P_over_P}
\frac{\left.\delta\specificsurface_{\rm ABCD}\right|_\philiq}
{\delta L/L}&\simeq&
\frac{\pi}{4}\,\frac{L\,H}{\Vbub}\\
\label{Eq:delta_specific_surface_area_EFG_over_delta_P_over_P}
\frac{\left.\delta\specificsurface_{\rm EFG}\right|_\philiq}
{\delta L/L}&\simeq&
\frac{L\,H}{\Vbub}
\eee
where it turns out that 
Eq.~(\ref{Eq:delta_specific_surface_area_EFG_over_delta_P_over_P})
for regimes E-G
results just from the change in perimeter,
{\em i.e.}, the $\delta L$ term
in Eq.~(\ref{Eq:delta_specific_surface_area_EFG}).
By contrast, for regimes A-D, the result 
of Eq.~(\ref{Eq:delta_specific_surface_area_ABCD_over_delta_P_over_P})
depends partly on the reduction in Plateau border radius
($\delta\Pbradius$ term)
in Eq.~(\ref{Eq:delta_specific_surface_area_ABCD})
caused by the deformation.

Let us now derive the shear modulus $G$
from the above equations.
In the case of a foam, 
the elastic energy per unit volume in the material
is given by the change in specific surface area
caused by some (small) shear strain $\Gamma$,
multiplied by surface tension:
\be
\label{Eq:shear_modulus_related_change_specific_surface_area}
\frac12\,G\,\Gamma^2
=\surfacetension\,\delta\specificsurface(\Gamma)
\ee
Now, because shear causes both deformation and rotation,
the effect of a strain $\Gamma$
on elongation is halved: $\lambda=1+\Gamma/2$.
Hence, Eq.~(\ref{Eq:relative_dry_perimeter_increase}) yields:
\be
\label{Eq:delta_P_fonction_Gamma}
\left.\frac{\delta L}{L}\right|_\philiq
\simeq\frac{\Gamma^2}{8}
\ee
Using Eqs.~(\ref{Eq:shear_modulus_related_change_specific_surface_area})
and~(\ref{Eq:delta_P_fonction_Gamma}),
the shear modulus in the pancake and in the floor tile regimes
can now be derived from 
Eqs.~(\ref{Eq:delta_specific_surface_area_ABCD_over_delta_P_over_P})
and~(\ref{Eq:delta_specific_surface_area_EFG_over_delta_P_over_P}):
\bee
\label{Eq:shear_modulus_ABCD}
G^{\rm ABCD}&\simeq&\frac{\pi}{16}
\,\frac{\surfacetension\,L\,H}{\Vbub}\\
\label{Eq:shear_modulus_EFG}
G^{\rm EFG}&\simeq&\frac{1}{4}
\,\frac{\surfacetension\,L\,H}{\Vbub}
\eee

The results of Eqs.~(\ref{Eq:shear_modulus_ABCD})
and~(\ref{Eq:shear_modulus_EFG}) above
are reported in Table~\ref{tab:geom_gg}.

\subsection{Change in osmotic pressure}

The change in bubble perimeter
also causes a change in osmotic pressure.

From Eqs.~(\ref{Eq:piosm_zz_ABCD}-\ref{Eq:piosm_pl_EFG}),
we obtain:
\bee
\frac{\delta\piosm^{zz\,\rm ABCD}}{1-\philiq}
&+&\frac{\piosm^{zz\,\rm ABCD}\,\delta\philiq}{(1-\philiq)^2}\nonumber\\
&&\simeq -\frac23\,\frac{\surfacetension}{\Pbradius^2}\,\delta\Pbradius
-\frac{\pi}{4}\,\frac{\surfacetension\,H}{\Vbub}\,\delta L
\label{Eq:delta_piosm_zz_ABCD_delta_P_delta_R}\\
%
\frac{\delta\piosm^{pl\,\rm ABCD}}{1-\philiq}
&+&\frac{\piosm^{pl\,\rm ABCD}\,\delta\philiq}{(1-\philiq)^2}\nonumber\\
&&\simeq -\frac23\,\frac{\surfacetension}{\Pbradius^2}\,\delta\Pbradius
-\frac{\pi}{8}\,\frac{\surfacetension\,H}{\Vbub}\,\delta L
\label{Eq:delta_piosm_pl_ABCD_delta_P_delta_R}\\
%
\frac{\delta\piosm^{zz\,\rm EFG}}{1-\philiq}
&+&\frac{\piosm^{zz\,\rm EFG}\,\delta\philiq}{(1-\philiq)^2}\nonumber\\
&&\simeq-\frac{\surfacetension}{\Pbradius^2}\,\delta\Pbradius
-\frac{\surfacetension\,H}{\Vbub}\,\delta L
\label{Eq:delta_piosm_zz_EFG_delta_P_delta_R}\\
%
\frac{\delta\piosm^{pl\,\rm EFG}}{1-\philiq}
&+&\frac{\piosm^{pl\,\rm EFG}\,\delta\philiq}{(1-\philiq)^2}\nonumber\\
&&\simeq-\frac{\surfacetension}{\Pbradius^2}\,\delta\Pbradius
-\frac{\surfacetension\,H}{2\,\Vbub}\,\delta L
\label{Eq:delta_piosm_pl_EFG_delta_P_delta_R}
\eee
where some terms have been neglected 
as explained in Paragraph~\ref{Sec:simplifications}.

At constant volume fraction ($\delta\philiq=0$)
and with $\philiq\ll 1$,
the above equations now yield,
using Eqs.~(\ref{Eq:delta_R_fonction_delta_P_ABCD})
and~(\ref{Eq:delta_R_fonction_delta_P_EFG}):
\bee
\frac{\delta\piosm^{zz\,\rm ABCD}}{\delta L/L}
&\simeq&
\frac{\frac{\surfacetension}{H}
-\frac{3\pi(4\sqrt{3}-2\pi)}{(4-\pi)}
\,\frac{\surfacetension\,\Pbradius^3}{\Vbub\,H}
}
{1+\frac{12(4\sqrt{3}-2\pi)}{4-\pi}\,\frac{\Pbradius^3}{H^2\,L}}
\label{Eq:delta_piosm_zz_ABCD}\\
\frac{\delta\piosm^{pl\,\rm ABCD}}{\delta L/L}
&\simeq&
\frac{\frac{\surfacetension}{H}
-\frac{3\pi(4\sqrt{3}-2\pi)}{2(4-\pi)}
\,\frac{\surfacetension\,\Pbradius^3}{\Vbub\,H}
}
{1+\frac{12(4\sqrt{3}-2\pi)}{4-\pi}\,\frac{\Pbradius^3}{H^2\,L}}
\label{Eq:delta_piosm_pl_ABCD}\\
\frac{\delta\piosm^{zz\,\rm EFG}}{\delta L/L}
&\simeq&
-\frac{\surfacetension\,H\,L}{\Vbub}
+\frac{\frac{\surfacetension}{2\,\Pbradius}}
{1+\frac{4\sqrt{3}-2\pi}{4-\pi}\,\frac{H}{L}}
\label{Eq:delta_piosm_zz_EFG}\\
\frac{\delta\piosm^{pl\,\rm EFG}}{\delta L/L}
&\simeq&
-\frac{\surfacetension\,H\,L}{2\,\Vbub}
+\frac{\frac{\surfacetension}{2\,\Pbradius}}
{1+\frac{4\sqrt{3}-2\pi}{4-\pi}\,\frac{H}{L}}
\label{Eq:delta_piosm_pl_EFG}
\eee
where, again, some terms have been neglected.
These results are reported in Table~\ref{tab:results}
in each asymptotic regime,
using the relations of Table~\ref{tab:geometrical_transitions}.
Note that both normal and in-plane dilatancy 
are negative not only in regime G as mentioned
in ref.~\cite{dilatancy_geometry_letter_2008}, 
but also in regime A.

\subsection{Elastic dilatancy in GG foams: shear or elongation, in-plane or normal}
\label{Sec:dilatancy_gg_foams}

There are of course two versions of dilatancy,
depending on whether the osmotic pressure is measured
in the plane of deformation or in the normal direction.
But the value of the coefficient 
defined by Eq.~(\ref{Eq:def_dilatancy_derivative})
depends on the deformation mode that is considered.
Hence, for an elongation by a factor  $\lambda=1+\epsilon$
or for a shear strain $\Gamma$,
the variation of the osmotic pressure
and the definition of the dilatancy coefficient will be:
\bee
\piosm\simeq\piosm^0+\frac{\epsilon^2}{2}
\left.\frac{\partial^2\piosm}{\partial\epsilon^2}\right|_\philiq
\hs
\dilatancy^{\rm el}=
\left.\frac{\partial^2\piosm}{\partial\epsilon^2}\right|_\philiq&&\\
\piosm\simeq\piosm^0+\frac{\Gamma^2}{2}
\left.\frac{\partial^2\piosm}{\partial\Gamma^2}\right|_\philiq
\hs
\dilatancy^{\rm sh}=
\left.\frac{\partial^2\piosm}{\partial\Gamma^2}\right|_\philiq&&
\eee
The dilatancy coefficient can thus be expressed as:
\bee
\dilatancy^{\rm el}&=&
\frac{\left.\delta\piosm(\epsilon)\right|_\philiq}
{\epsilon^2/2}\\
\dilatancy^{\rm sh}&=&
\frac{\left.\delta\piosm(\Gamma)\right|_\philiq}
{\Gamma^2/2}
\eee
Now, for elongation, the usual definition of a deformation $\epsilon$
is that the material is elongated by a factor $\lambda=1+\epsilon$.
Hence, Eq.~(\ref{Eq:relative_dry_perimeter_increase}) yields:
\be
\label{Eq:delta_P_fonction_epsilon}
\left.\frac{\delta L}{L}\right|_\philiq
\simeq\frac{\epsilon^2}{2}
\ee
As a result of
Eqs.~(\ref{Eq:delta_P_fonction_Gamma})
and~(\ref{Eq:delta_P_fonction_epsilon}),
the dilatancy coefficients 
for elongation and shear deformation modes 
can be expressed as:
\bee
\label{Eq:dilatancy_elongation}
\dilatancy^{\rm el}
&\simeq&\frac{\left.\delta\piosm\right|_\philiq}
{\left.(\delta L/L)\right|_\philiq}\\
\label{Eq:dilatancy_shear}
\dilatancy^{\rm sh}&=&\frac14\,\dilatancy^{\rm el}
\eee

From Eq.~(\ref{Eq:dilatancy_elongation}
)
and Eqs.~(\ref{Eq:delta_piosm_zz_ABCD}-\ref{Eq:delta_piosm_pl_EFG}),
the elongation normal and in-plane dilatancies
are indicated in Table~\ref{tab:results}
for each regime specified
in Table~\ref{tab:geometrical_transitions}.

\subsection{Change in volume fraction}

We shall now calculate the change in liquid volume fraction $\philiq$
that results from the foam deformation (change in perimeter)
under constant (normal or in-plane) osmotic pressure.

To do this, we set $\delta\piosm=0$
in Eqs.~(\ref{Eq:delta_piosm_zz_ABCD_delta_P_delta_R}-\ref{Eq:delta_piosm_pl_EFG_delta_P_delta_R})
and we eliminate $\delta\Pbradius$
between each of these equations
and Eq.~(\ref{Eq:delta_philiq_ABCD_function_delta_R_and_delta_P})
or~(\ref{Eq:delta_philiq_EFG_function_delta_R_and_delta_P}).
We thus obtain:
\bee
\label{Eq:delta_philiq_over_delta_P_over_P_ABCD_zz}
\left.\frac{\delta\philiq^{\rm ABCD}}{\delta L/L}\right|_{\piosm^{zz}}
&\simeq&\frac{\frac{4-\pi}{8}\,\frac{H^2\,L}{\Vbub}
-\frac{3\pi(4\sqrt{3}-2\pi)}{8}\,\frac{\Pbradius^3\,H^2\,L}{\Vbub^2}}
{1+3(4\sqrt{3}-2\pi)\,\frac{\Pbradius^3}{\Vbub}}
\\
\label{Eq:delta_philiq_over_delta_P_over_P_ABCD_pl}
\left.\frac{\delta\philiq^{\rm ABCD}}{\delta L/L}\right|_{\piosm^{pl}}
&\simeq&\frac{4-\pi}{8}\,\frac{H^2\,L}{\Vbub}\nonumber\\
&&-\frac{3\pi(4\sqrt{3}-2\pi)}{16}\,\frac{\Pbradius^3\,H^2\,L}{\Vbub^2}
\\
\label{Eq:delta_philiq_over_delta_P_over_P_EFG_zz}
\left.\frac{\delta\philiq^{\rm EFG}}{\delta L/L}\right|_{\piosm^{zz}}
&\simeq&
\frac{4-\pi}{2}\,\frac{\Pbradius^2\,L}{\Vbub}\nonumber\\
&&-(4\sqrt{3}-2\pi)\,\frac{\Pbradius^3\,H^2\,L}{\Vbub^2}
\\
\label{Eq:delta_philiq_over_delta_P_over_P_EFG_pl}
\left.\frac{\delta\philiq^{\rm EFG}}{\delta L/L}\right|_{\piosm^{pl}}
&\simeq&\frac{4-\pi}{2}\,\frac{\Pbradius^2\,L}{\Vbub}\nonumber\\
&&-(2\sqrt{3}-\pi)\,\frac{\Pbradius^3\,H^2\,L}{\Vbub^2}
\eee

The relative change in liquid volume fraction
can then be obtained 
from Eqs.~(\ref{Eq:philiq_ABCD_Vbub}-\ref{Eq:philiq_EFG_Vbub}):
\bee
\label{Eq:delta_philiq_over_philiq_over_delta_P_over_P_ABCD_zz}
&\left.\frac{\delta\philiq/\philiq}
{\delta L/L}\right|_{\piosm^{zz}}^{\rm ABCD}
\simeq\frac{1-\frac{3\pi(4\sqrt{3}-2\pi)}{4-\pi}
\,\frac{\Pbradius^3}{\Vbub}}
{\left[1+3(4\sqrt{3}-2\pi)\,\frac{\Pbradius^3}{\Vbub}\right]
\,\left[1+\frac{4(4\sqrt{3}-2\pi)}{4-\pi}
\,\frac{\Pbradius^2}{L\,H}\right]}
&\\
\label{Eq:delta_philiq_over_philiq_over_delta_P_over_P_ABCD_pl}
&\left.\frac{\delta\philiq/\philiq}
{\delta L/L}\right|_{\piosm^{pl}}^{\rm ABCD}
\simeq\frac{1-\frac{3\pi(4\sqrt{3}-2\pi)}{2(4-\pi)}\,\frac{\Pbradius^3}{\Vbub}}
{1+\frac{4(4\sqrt{3}-2\pi)}{4-\pi}\,\frac{\Pbradius^2}{L\,H}}
&\\
\label{Eq:delta_philiq_over_philiq_over_delta_P_over_P_EFG_zz}
&\left.\frac{\delta\philiq/\philiq}
{\delta L/L}\right|_{\piosm^{zz}}^{\rm EFG}
\simeq\frac{
1-\frac{2(4\sqrt{3}-2\pi)}{4-\pi}\,\frac{\Pbradius\,H^2}{\Vbub}
}
{1+\frac{4\sqrt{3}-2\pi}{4-\pi}\,\frac{H}{L}}
&\\
\label{Eq:delta_philiq_over_philiq_over_delta_P_over_P_EFG_pl}
&\left.\frac{\delta\philiq/\philiq}
{\delta L/L}\right|_{\piosm^{pl}}^{\rm EFG}
\simeq\frac{
1-\frac{4\sqrt{3}-2\pi}{4-\pi}\,\frac{\Pbradius\,H^2}{\Vbub}
}
{1+\frac{4\sqrt{3}-2\pi}{4-\pi}\,\frac{H}{L}}
&
\eee

\bibliography{article_dilatancy_2010}

\begin{thebibliography}{15}
\expandafter\ifx\csname natexlab\endcsname\relax\def\natexlab#1{#1}\fi
\expandafter\ifx\csname bibnamefont\endcsname\relax
  \def\bibnamefont#1{#1}\fi
\expandafter\ifx\csname bibfnamefont\endcsname\relax
  \def\bibfnamefont#1{#1}\fi
\expandafter\ifx\csname citenamefont\endcsname\relax
  \def\citenamefont#1{#1}\fi
\expandafter\ifx\csname url\endcsname\relax
  \def\url#1{\texttt{#1}}\fi
\expandafter\ifx\csname urlprefix\endcsname\relax\def\urlprefix{URL }\fi
\providecommand{\bibinfo}[2]{#2}
\providecommand{\eprint}[2][]{\url{#2}}

\bibitem[{\citenamefont{Weaire and Hutzler}(1999)}]{weaire_hutzler_1999_book}
\bibinfo{author}{\bibfnamefont{D.}~\bibnamefont{Weaire}} \bibnamefont{and}
  \bibinfo{author}{\bibfnamefont{S.}~\bibnamefont{Hutzler}},
  \emph{\bibinfo{title}{The Physics of Foams}} (\bibinfo{publisher}{Oxford
  University Press}, \bibinfo{year}{1999}).

\bibitem[{\citenamefont{Cantat et~al.}(2010)\citenamefont{Cantat, Cohen-Addad,
  Elias, Graner, H{\\"o}hler, Pitois, Rouyer, and
  Saint-Jalmes}}]{french_book_belin_2010}
\bibinfo{author}{\bibfnamefont{I.}~\bibnamefont{Cantat}},
  \bibinfo{author}{\bibfnamefont{S.}~\bibnamefont{Cohen-Addad}},
  \bibinfo{author}{\bibfnamefont{F.}~\bibnamefont{Elias}},
  \bibinfo{author}{\bibfnamefont{F.}~\bibnamefont{Graner}},
  \bibinfo{author}{\bibfnamefont{R.}~\bibnamefont{H{\\"o}hler}},
  \bibinfo{author}{\bibfnamefont{O.}~\bibnamefont{Pitois}},
  \bibinfo{author}{\bibfnamefont{F.}~\bibnamefont{Rouyer}}, \bibnamefont{and}
  \bibinfo{author}{\bibfnamefont{A.}~\bibnamefont{Saint-Jalmes}},
  \emph{\bibinfo{title}{Les mousses - structure et dynamique}}
  (\bibinfo{publisher}{Belin}, \bibinfo{address}{Paris}, \bibinfo{year}{2010}).

\bibitem[{\citenamefont{Reynolds}(1985)}]{reynolds_centenary_1985_469}
\bibinfo{author}{\bibfnamefont{O.}~\bibnamefont{Reynolds}},
  \bibinfo{journal}{Philos. Mag.} \textbf{\bibinfo{volume}{20}},
  \bibinfo{pages}{469} (\bibinfo{year}{1985}).

\bibitem[{\citenamefont{Bagnold}(1941)}]{bagnold_1941}
\bibinfo{author}{\bibfnamefont{R.}~\bibnamefont{Bagnold}},
  \emph{\bibinfo{title}{The physics of blown sand and desert dunes}}
  (\bibinfo{publisher}{Chapman and Hall}, \bibinfo{address}{London},
  \bibinfo{year}{1941}).

\bibitem[{\citenamefont{Marze et~al.}(2005)\citenamefont{Marze, Saint-Jalmes,
  and Langevin}}]{marze_2005_121}
\bibinfo{author}{\bibfnamefont{S.}~\bibnamefont{Marze}},
  \bibinfo{author}{\bibfnamefont{A.}~\bibnamefont{Saint-Jalmes}},
  \bibnamefont{and} \bibinfo{author}{\bibfnamefont{D.}~\bibnamefont{Langevin}},
  \bibinfo{journal}{Colloids and Surface A} \textbf{\bibinfo{volume}{263}},
  \bibinfo{pages}{121} (\bibinfo{year}{2005}).

\bibitem[{\citenamefont{Weaire and Hutzler}(2003)}]{weaire_2003_2747}
\bibinfo{author}{\bibfnamefont{D.}~\bibnamefont{Weaire}} \bibnamefont{and}
  \bibinfo{author}{\bibfnamefont{S.}~\bibnamefont{Hutzler}},
  \bibinfo{journal}{Phil. Mag.} \textbf{\bibinfo{volume}{83}},
  \bibinfo{pages}{2747} (\bibinfo{year}{2003}).

\bibitem[{\citenamefont{Rioual et~al.}(2005)\citenamefont{Rioual, Hutzler, and
  Weaire}}]{rioual_2005_117}
\bibinfo{author}{\bibfnamefont{F.}~\bibnamefont{Rioual}},
  \bibinfo{author}{\bibfnamefont{S.}~\bibnamefont{Hutzler}}, \bibnamefont{and}
  \bibinfo{author}{\bibfnamefont{D.}~\bibnamefont{Weaire}},
  \bibinfo{journal}{Coll. Surf. A} \textbf{\bibinfo{volume}{263}},
  \bibinfo{pages}{117} (\bibinfo{year}{2005}).

\bibitem[{\citenamefont{Rognon et~al.}(2010{\natexlab{a}})\citenamefont{Rognon,
  Molino, and Gay}}]{dilatancy_geometry_letter_2008}
\bibinfo{author}{\bibfnamefont{P.}~\bibnamefont{Rognon}},
  \bibinfo{author}{\bibfnamefont{F.}~\bibnamefont{Molino}}, \bibnamefont{and}
  \bibinfo{author}{\bibfnamefont{C.}~\bibnamefont{Gay}}, \bibinfo{journal}{EPL
  (Europhysics Letters)} \textbf{\bibinfo{volume}{90}}, \bibinfo{pages}{38001}
  (\bibinfo{year}{2010}{\natexlab{a}}).

\bibitem[{\citenamefont{Vaz and Cox}(2005)}]{Vaz_Cox_2005_415}
\bibinfo{author}{\bibfnamefont{M.}~\bibnamefont{Vaz}} \bibnamefont{and}
  \bibinfo{author}{\bibfnamefont{S.}~\bibnamefont{Cox}},
  \bibinfo{journal}{Philosophical Magazine Letters}
  \textbf{\bibinfo{volume}{85}}, \bibinfo{pages}{415} (\bibinfo{year}{2005}).

\bibitem[{\citenamefont{Rognon et~al.}(2010{\natexlab{b}})\citenamefont{Rognon,
  Gay, Reinelt, and Molino}}]{bulimia_2010}
\bibinfo{author}{\bibfnamefont{P.}~\bibnamefont{Rognon}},
  \bibinfo{author}{\bibfnamefont{C.}~\bibnamefont{Gay}},
  \bibinfo{author}{\bibfnamefont{D.}~\bibnamefont{Reinelt}}, \bibnamefont{and}
  \bibinfo{author}{\bibfnamefont{F.}~\bibnamefont{Molino}},
  \bibinfo{journal}{subm. to Eur. Phys. J. E}
  (\bibinfo{year}{2010}{\natexlab{b}}),
  \eprint{http://hal.archives-ouvertes.fr/hal-00361004/fr/}.

\bibitem[{\citenamefont{Cox et~al.}(2002)\citenamefont{Cox, Weaire, and
  Vaz}}]{cox_epje2002_311}
\bibinfo{author}{\bibfnamefont{S.}~\bibnamefont{Cox}},
  \bibinfo{author}{\bibfnamefont{D.}~\bibnamefont{Weaire}}, \bibnamefont{and}
  \bibinfo{author}{\bibfnamefont{M.}~\bibnamefont{Vaz}}, \bibinfo{journal}{The
  European Physical Journal E: Soft Matter and Biological Physics}
  \textbf{\bibinfo{volume}{7}}, \bibinfo{pages}{311} (\bibinfo{year}{2002}),
  ISSN \bibinfo{issn}{1292-8941},
  \urlprefix\url{http://dx.doi.org/10.1140/epje/i2001-10099-1}.

\bibitem[{\citenamefont{Cox and Janiaud}(2008)}]{Cox_Janiaud_2008_693}
\bibinfo{author}{\bibfnamefont{S.}~\bibnamefont{Cox}} \bibnamefont{and}
  \bibinfo{author}{\bibfnamefont{E.}~\bibnamefont{Janiaud}},
  \bibinfo{journal}{Philosophical Magazine Letters}
  \textbf{\bibinfo{volume}{88}}, \bibinfo{pages}{693} (\bibinfo{year}{2008}).

\bibitem[{\citenamefont{Graner et~al.}(2008)\citenamefont{Graner, Dollet,
  Raufaste, and Marmottant}}]{graner_2008_369}
\bibinfo{author}{\bibfnamefont{F.}~\bibnamefont{Graner}},
  \bibinfo{author}{\bibfnamefont{B.}~\bibnamefont{Dollet}},
  \bibinfo{author}{\bibfnamefont{C.}~\bibnamefont{Raufaste}}, \bibnamefont{and}
  \bibinfo{author}{\bibfnamefont{P.}~\bibnamefont{Marmottant}},
  \bibinfo{journal}{Eur. Phys. J. E} \textbf{\bibinfo{volume}{25}},
  \bibinfo{pages}{349} (\bibinfo{year}{2008}), \eprint{arXiv:0708.3193
  [cond-mat.soft]}.

\bibitem[{\citenamefont{Aubouy et~al.}(2003)\citenamefont{Aubouy, Jiang,
  Glazier, and Graner}}]{aubouy_2003_67}
\bibinfo{author}{\bibfnamefont{M.}~\bibnamefont{Aubouy}},
  \bibinfo{author}{\bibfnamefont{Y.}~\bibnamefont{Jiang}},
  \bibinfo{author}{\bibfnamefont{J.}~\bibnamefont{Glazier}}, \bibnamefont{and}
  \bibinfo{author}{\bibfnamefont{F.}~\bibnamefont{Graner}},
  \bibinfo{journal}{Granular Matter} \textbf{\bibinfo{volume}{5}},
  \bibinfo{pages}{67} (\bibinfo{year}{2003}).

\bibitem[{\citenamefont{Asipauskas et~al.}(2003)\citenamefont{Asipauskas,
  Aubouy, Glazier, Graner, and Jiang}}]{asipauskas_2003_71}
\bibinfo{author}{\bibfnamefont{M.}~\bibnamefont{Asipauskas}},
  \bibinfo{author}{\bibfnamefont{M.}~\bibnamefont{Aubouy}},
  \bibinfo{author}{\bibfnamefont{J.}~\bibnamefont{Glazier}},
  \bibinfo{author}{\bibfnamefont{F.}~\bibnamefont{Graner}}, \bibnamefont{and}
  \bibinfo{author}{\bibfnamefont{Y.}~\bibnamefont{Jiang}},
  \bibinfo{journal}{Granular Matter} \textbf{\bibinfo{volume}{5}},
  \bibinfo{pages}{71} (\bibinfo{year}{2003}).

\end{thebibliography}
\end{document}